\documentclass[letter]{aa}  
\usepackage{graphicx}
\usepackage{psfig}
\usepackage{txfonts}
\begin{document}
   \title{On the optical counterpart
of NGC~300 X-1 and the global 
Wolf-Rayet content of NGC~300\thanks{Based on 
observations made with ESO
Telescopes at the Paranal Observatory  under programme ID 278.D-5019}}

   \author{Paul A. Crowther\inst{1}, S. Carpano\inst{2}, 
L.J. Hadfield\inst{1}, A.M.T. Pollock\inst{2}}

   \offprints{Paul Crowther (Paul.Crowther@sheffield.ac.uk)}

   \institute{Department of Physics \& Astronomy, University of Sheffield,
Hicks Building, Hounsfield Rd, Sheffield, S3 7RH, United Kingdom \and
XMM-Newton Science Operations Center, ESAC, 28080, Madrid, Spain}

%             }

   \date{}

 \abstract
{
Surveys of Wolf-Rayet (WR) populations in nearby galaxies provide tests of evolutionary
models plus Type Ib/c supernova progenitors. This spectroscopic study
complements the recent imaging survey of the spiral galaxy NGC~300 by Schild et al.}
{Revisions to the known WR content of 
NGC~300 are presented.
We investigate the WR nature of candidate \#41 from Schild et al. which is
spatially coincident with the bright X-ray point source NGC~300 X-1.}
{VLT/FORS2 multi-object spectroscopy of WR candidates in NGC~300 is 
obtained.}
{We establish an early-type WN nature of \#41, i.e. similar
to the optical counterpart of IC~10 X-1, which closely resembles NGC~300 X-1. 
We confirm 9 new WR stars, bringing the current WR census 
of the inner disk to 31, with N(WC)/N(WN)$\sim$0.9.}
{If \#41 is the optical counterpart for NGC~300 X-1, we estimate a WR mass of
38 M$_{\odot}$ based upon ground-based photometry, from which a 
black hole mass of $\geq$ 10 M$_{\odot}$ results from the 32.8~hr period 
of
the system and WR wind velocity of 1250 km\,s$^{-1}$. We estimate an 95\% completeness 
among WC stars and 70\% among WN stars, such that the total WR content is
$\sim$40, with N(WC)/N(WN)$\sim$0.7. From the H$\alpha$-derived star formation 
rate of the inner galaxy, we infer N(WR)/N(O)$\sim$0.04.
} 
% 5 {} token are mandatory
 

\authorrunning{P.A. Crowther et al.}
\titlerunning{WR content of NGC~300}

   \maketitle
%
%________________________________________________________________

\section{Introduction}

Massive stars dominate the feedback to the local interstellar medium
(ISM)  in star-forming galaxies via their stellar winds and ultimate
death as core-collapse supernovae (SNe). In particular, Wolf-Rayet (WR)  
stars typically have wind densities an order of magnitude higher than
massive O stars. They contribute to the chemical enrichment of
galaxies, they are the prime candidates for the
immediate progenitors of long, soft Gamma Ray Bursts (GRBs), and they 
provide a signature of high-mass star formation in galaxies (Crowther 
2007).

WR stars exhibit a unique broad emission line spectral appearance,
for which He\,{\sc ii} $\lambda$4686 (WN subtypes) and C\,{\sc 
iii} $\lambda$4647-51, C\,{\sc iv} $\lambda$5801-12 (WC subtypes) 
provide the basis for their detection in external galaxies.
Narrow-band filter techniques have been developed by 
Moffat, Seggewiss \& Shara (1985) and Massey, Armandroff \& 
Conti (1986) and applied to Local Group galaxies.
Attempts have been made with 4m telescopes to extend the
technique beyond the Local Group, although this proved to be
challenging for the southern spiral NGC~300 
(Schild \& Testor 1991) at a distance of 1.88~Mpc (Gieren et al. 2005).

The advent of efficient, multi-object spectrographs at 8m telescopes,
such as FORS1/2 at the  Very Large Telescope, has permitted surveys of WR populations in 
galaxies at distances of several Mpc. 
Such narrow-band surveys -- NGC~300 (Schild et al. 2003),
M~83 (Hadfield et al. 2005), NGC~1313 (Hadfield \& Crowther 2007)
-- complement ongoing  
broad-band surveys for SN progenitors (e.g. Smartt et al. 2004). 
Indeed, WN and WC stars 
are believed to be the immediate progenitors for a subset of 
Type  Ib (H-poor) and Type Ic (H, He-poor) SN. Progenitors
of Type Ib/c SN remain to be established, for which broad-band 
surveys would fail to confirm the emission-line (WR) nature.

It is well established that the absolute number of WR stars and their 
subtype distribution are metallicity dependent. N(WR)/N(O)$\sim$0.15 in 
the metal-rich Solar neighbourhood, yet N(WR)/N(O)$\sim$0.01 in 
the metal-deficient SMC (Crowther 2007). 
This observational dependence results from the easier removal of
progenitor O star hydrogen-envelopes at high metallicity, since
O-type stars possess winds that are driven by metallic lines 
(Mokiem et al. 2007). Consequently, the WR  
content of star forming galaxies spanning a range of metallicities provide 
useful test cases for evolutionary and population synthesis models (Meynet 
\& Maeder 2005).

In addition, high spatial resolution X-ray surveys of nearby galaxies with 
{\it Chandra} and {\it XMM-Newton} permit the identification of exotic 
binaries. Carpano et al. (2007a) established that the brightest 
X-ray point source in NGC~300 X-1 was spatially coincident with the 
candidate WR star \#41 from Schild et al. (2003). Systems involving WR 
stars and a close neutron star (NS) or black hole (BH) companion represent 
a natural, albeit rare, end state for close binary evolution, for which 
only Cyg X-3  in the Milky Way and IC~10 X-1 are confirmed 
to date (Lommen et al. 2005). 
Establishing the nature of \#41  is the 
primary motivation for the present Letter, together with 
a more comprehensive study of the WR content of NGC~300.

\section{Observations}

We used the ESO Very Large Telescope and Focal Reducer/Low Dispersion 
Spectrograph \#2 (FORS2) in multi-object spectroscopy (MOS) mode on 12 Jan 
2007 using three 800~s exposures with the 300V grism, centred at 
$\lambda$=5900\AA. 16 WR candidates from Schild et al. (2003) were 
observed simultaneously using MOS with  1.0~arcsec slits, 
of which 4 sources had previously been spectroscopically confirmed. 
Seeing conditions were 1.1--1.2 arcsec at relatively high 
airmass of 1.7--2.0.

\begin{figure} 
%\centerline{\psfig{figure=newwr2.eps,width=7.5cm,angle=0}}
\centerline{\psfig{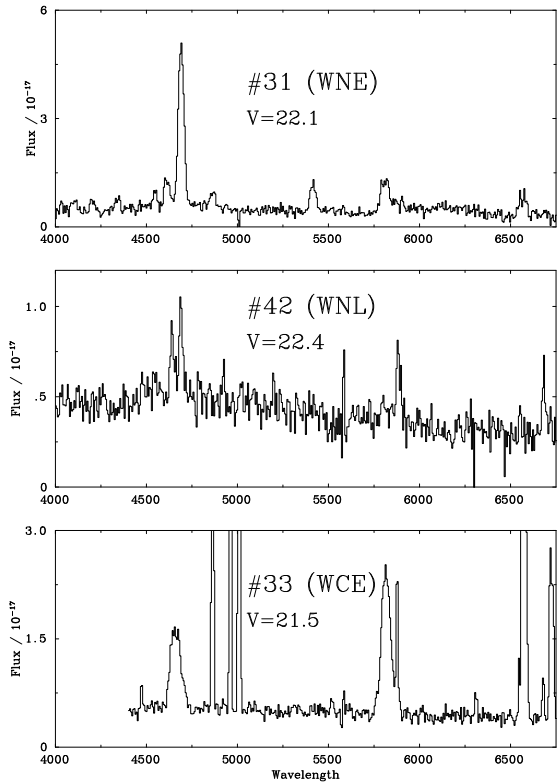}}
\caption{VLT/FORS2 spectroscopy of three new WR stars.
Narrow emission lines in the spectrum of \#33 are from
an associated H\,{\sc ii} region.}\label{newwr2}
\end{figure}

A standard extraction was performed using {\sc iraf}, with wavelength and 
flux calibration carried out using {\sc figaro}. The spectral resolution, 
as measured from comparison arc lines was $\sim$10\AA. The flux standard 
LTT~1788 provided a relative flux calibration. Flux calibration of \#9, \#24,
\#30 and \#40, in common with spectroscopy of 
Schild et al. (2003), were found to be in excellent agreement.
Absolute flux calibration 
was achieved by comparing the photometry of an individual object in the 
$\lambda$4684 filter (from Schild et al. 2003) with a synthetic 
photometric measurement, determined by convolving our spectroscopy with a 
suitable Gaussian filter and zero-point. 

\begin{table}[htbp]
\caption[]{List of Wolf-Rayet candidates from Schild et al. (2003) spectroscopically
observed with VLT/FORS2, $\Delta m = m_{\rm 4781} - m_{\rm 4686}$.
 Candidate \#41 is spatially coincident with NGC~300 X-1.}
\label{table1}
\begin{tabular}{r@{\hspace{2.5mm}}c@{\hspace{2.5mm}}c
@{\hspace{2mm}}r@{\hspace{2mm}}c@{\hspace{-1mm}}r@{\hspace{1mm}}c@{\hspace{-1mm}}c}
\noalign{\hrule\smallskip}
ID & $m_{\rm 4781}$ & $\Delta m$ & \multicolumn{2}{c}{C\,{\sc 
iii} 4650/He\,{\sc ii} 4686} & \multicolumn{2}{c}{C\,{\sc iv} 5808} &
Sp.Type \\
 & mag & mag & $W_{\lambda}$ (\AA) & FWHM (\AA) & $W_{\lambda}$ (\AA) & 
FWHM (\AA) \\
\noalign{\smallskip\hrule\smallskip}
3 & 23.13 & 1.25 & 110$\pm$6 & 24$\pm$1 & -- & -- & WNE \\
5 & 22.61 & 2.09 & 44$\pm$1 & 40$\pm$1 &28$\pm$2& 35$\pm$2  & WCE \\
8 & 20.46 & 0.22 &  3$\pm$1 & 17$\pm$3 & -- & -- & WNL \\
12 & 22.44 & 1.57 & 148$\pm$5 & 30$\pm$1 & -- & -- & WNE \\
20 & 18.65 & 0.11 & -- & -- & -- & -- &  Not WR\\
23 & 19.80 & 0.08 & -- & -- & -- & -- &  Not WC \\
27 & 20.54 & 0.15 & 3$\pm$1 & 10$\pm$2 & -- & -- & Of \\
31 & 22.42 & 1.68 & 216$\pm$7 & 34$\pm$1 & -- & -- & WNE  \\
33 & 22.30 & 1.02 &  149$\pm$6 & 66$\pm$2 & 266$\pm$5 & 58$\pm$1  & WCE \\
41 & 22.59 & 0.95 & 53$\pm$3 & 19$\pm$1 & -- & -- & WNE \\
42 & 22.58 & 0.56 & 17$\pm$3 & 19$\pm$4 & -- & -- & WNL  \\
52 & 22.66 & 0.50 & 397$\pm$18 & 74$\pm$3 & 590$\pm$20  & 83$\pm$3 &WCE \\
\noalign{\smallskip\hrule\smallskip}
\end{tabular}
\end{table}

\section{Wolf-Rayet content of NGC~300}

Of the 12 new WR candidates we can confirm 9 cases, including \#41 
(NGC~300 X-1), for which we defer a detailed discussion until the next 
section. A summary of our new VLT/FORS2 spectroscopic detections is presented in
Table~\ref{table1}, with  three new WR stars shown in Fig.~\ref{newwr2}. 
Of the remaining candidates, \#20 was 
established as a foreground late-type star, \#27  was identified
as an Of star with weak, narrow He\,{\sc ii} $\lambda$4686 emission 
(\#8 possessed weak broad emission). Candidate \#23 did not include the critical 
$\lambda$4686 region, so a WC nature was excluded.

\begin{figure} 
%\centerline{\psfig{figure=excess_new.eps,width=8.5cm,angle=0}}
\centerline{\psfig{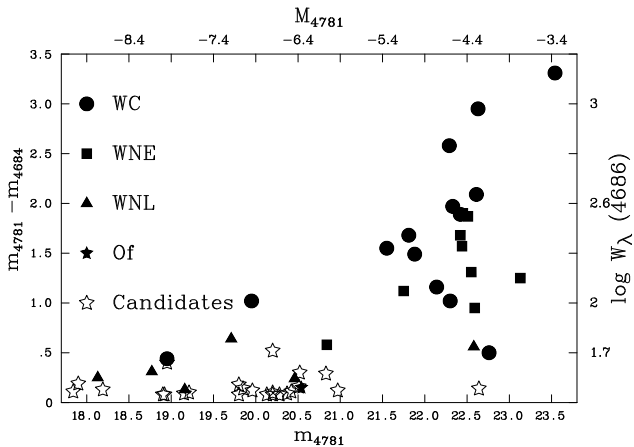}}
\caption{Comparison between the $\lambda4781$ continuum 
magnitudes of WR candidates in NGC~300 and the $\lambda$4684 excess
(stars), updated from Schild et al. (2003). 
Spectroscopically confirmed WR  stars are presented -- see
key. {\it Approximate} absolute magnitudes and line 
equivalent widths are indicated. Three WR candidates,
\#1 (WC4), \#37 and \#51, for which 
reliable photometry was not feasible  are omitted
from this figure.}\label{fig1}
\end{figure}

For the inner disk of NGC~300, our results decrease the observed
N(WC)/N(WN) ratio from 1.2 (12/10) to $\sim$0.9 (15/16). 
Our results provide
WR statistics for 30 sources versus 21 in Schild et al. (2003),
recalling that \#11 hosts both WN and WC stars. At face value, our
study is merely an incremental step, since sources are drawn
from a sample of 58 candidates within a de-projected radius of 5~arcmin (3~kpc). 
However, excluding \#20 and \#23, 
Fig.~\ref{fig1} shows that there are only 5 additional
candidates with a photometric excess greater than 0.3 mag.
Excesses of the remaining 21 cases are as little as 0.08 mag. 

\begin{figure} 
%\centerline{\psfig{figure=WC_WN_mono.eps,width=7cm,angle=0}}
\centerline{\psfig{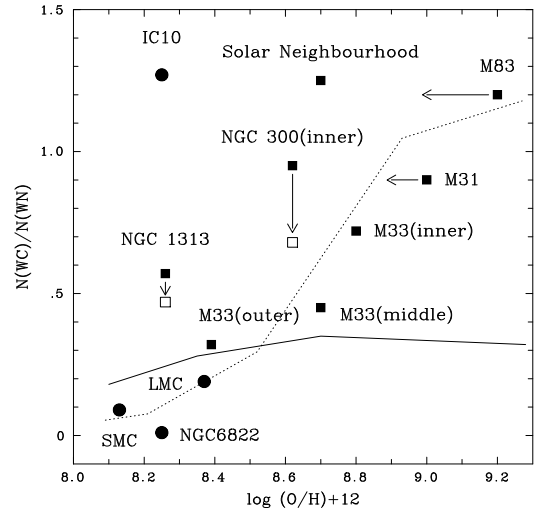}}
\caption{Spectroscopically confirmed N(WC)/N(WN) ratio for 
NGC~300, together with other nearby spiral (filled squares) and irregular 
(filled circles) galaxies from Crowther (2007) and references therein, 
plus Hadfield \& Crowther (2007) for NGC~1313. Open symbols for
NGC~300 and NGC~1313 are corrected for candidate WR sources,
as outlined in the text. Predictions from Meynet \& Maeder (2005, solid) 
and Eldridge \& Vink (2006, dotted) are also shown.}\label{WC_WN}
\end{figure}

Sources with large photometric excesses are likely to be WC stars
on the basis of their intrinsically stronger emission lines (cf. Fig.~\ref{newwr2}).
Indeed, all 4 spectroscopically
confirmed WR stars with photometric excesses greater than 2 mag are
WC stars (recall Fig.~\ref{fig1}). Sources with small photometric excesses will either be 
foreground late-type stars, Of stars or genuine WN stars. 
Assuming that \#37 is a WC star, with WN subtypes for all other candidates in excess of 0.15 mag,
and otherwise either an Of star or foreground late-type star. 
Consequently, the true WR 
content of the inner disk of NGC~300 is likely to increase from 31 to $\sim$40, with a subtype ratio closer to 
N(WC)/N(WN) $\sim$ 0.7. We compare the observed (filled symbols) and 
anticipated (open symbols) WR subtype distributions in NGC~300 and
NGC~1313 (Hadfield \& Crowther 2007) with spectroscopic results for
other nearby galaxies in  Figure~\ref{WC_WN}, together with evolutionary 
predictions from Meynet \& Maeder
(2005) and Eldridge \& Vink (2006). 
High N(WC)/N(WN) ratios for metal-rich
galaxies support the latter models, which allow for 
metallicity-dependent WR winds
(Crowther 2007).

The observed WC subtype distribution of relatively metal-poor galaxies, with
$\log$ (O/H) + 12 $\leq$ 8.5, is known to be dominated 
by early-type WC stars, e.g. such as the LMC and IC~10 (e.g. Crowther et al. 2003). 
In  contrast, more metal-rich galaxies with $\log$(O/H) 
+ 12 $\geq$ 8.5 will possess a fraction of late-type WC stars, 
as is the case for M\,33, M\,31 and the Milky Way. In the extreme
metal-rich environment of M\,83, with $\log$(O/H) + 12 $\sim$9, 
late subtypes dominate the WC population (Hadfield et al. 2005). 
Of the 15 known WC stars within the inner disk of 
NGC~300, all are early-types, i.e. the WC population is more in common
with other metal-poor galaxies.
However, the average metallicities of four H\,{\sc ii} regions 
(within $\rho/\rho_{0} \leq 0.5$) from Pagel et al. (1979) and Webster \& 
Smith (1983) is $\log$(O/H) + 12 = 8.62, close to the currently adopted 
Solar oxygen abundance of 8.66 (Asplund et al. 2004). 

The O star content of NGC~300 can most readily be obtained from 
H$\alpha$ imaging. Deharveng et al. (1988) provide a catalogue
of H\,{\sc ii} regions, including H$\alpha$ fluxes, the total output of which
is 1.3 $\times 10^{-11}$ erg\,s$^{-1}$\,cm$^{-2}$. Correction for
a typical extinction of $c(H\beta)$ = 0.21 from Webster \& Smith
(1983) implies a H$\alpha$ luminosity of 7.8$\times 10^{39}$ 
erg\,s$^{-1}$ adopting the Gieren et al. (2005) distance to NGC~300.
A global star formation rate (SFR) of 0.06 M$_{\odot}$ yr$^{-1}$ is 
implied for NGC~300, adopting the Kennicutt (1998) calibration. A
somewhat higher H$\alpha$ SFR of 0.11 M$_{\odot}$ yr$^{-1}$ was presented 
by Helou et al. (2004), adjusted for the Gieren et al. (2005) distance.
For our study, we adopt an intermediate global SFR of 0.085 M$_{\odot}$ 
yr$^{-1}$ -- a factor of two higher than the Small Magellanic Cloud (SMC,
Kennicutt 
et al. 1995) --  from which a Lyman continuum ionizing output of 8$\times 
10^{51}$ is inferred. From Kennicutt (1998) this is equal to 1,400 O7\,V 
equivalents based upon an O7\,V ionizing output of 10$^{48.75}$ (Martins
et al. 2005). 

Of course for a realistic N(WR)/N(O) ratio, we need to consider that
 Schild et al. (2003) surveyed the inner galaxy of 
NGC~300 (recall their fig.~1). From Deharveng et al. (1988) $\sim$78\% of 
the H$\alpha$ flux lies within this region, i.e. 1,100 equivalent 
O7\,V stars. As such, we obtain N(WR)/N(O)$\sim$0.04 on the basis of 
$\sim$40 WR stars. This is intermediate between that observed in the low 
metallicity SMC 
for which N(WR)/N(O)$\sim$0.01 and the Solar neighbourhood 
for which N(WR)/N(O)$\sim$0.15 (Crowther 2007). Again, the N(WR)/N(O) 
distribution suggests a metallicity that is $0.1-0.2$ dex more metal-poor than 
is suggested by Pagel et al. (1979) and Webster \& Smith (1983). Modern
nebular abundance studies of NGC~300 are underway (Bresolin, priv. comm.),
which will permit tests of our assertion that the metallicity of NGC~300 is
sub-solar.

\section{NGC~300 X-1}

One natural, albeit rare, product of close massive binary evolution 
involves a Wolf-Rayet star plus a compact (NS or BH) companion, if the 
system survives the core-collapse SN. Until recently, Cyg X-3 in 
the Milky Way, with a period of 4.8~hr and X-ray luminosity of 10$^{38}$ 
erg\,s$^{-1}$, was the only known example of an X-ray bright 
WR plus compact companion system
(van Kerkwijk 1993). Lommen et al. (2005) discuss the possible 
origins of  Cyg X-3, namely the accretion disk associated with a BH, fed 
by a massive WR star, or a NS fed through Roche lobe overflow by a lower 
mass helium star.

Bauer \& Brandt (2004) identified IC~10 X-1 as the first candidate 
extragalactic WR plus compact companion due to its spatial coincidence 
with the early-type WN star [MAC92]-17A from Crowther et al. (2003)
and Clark \& Crowther (2004). 
Recently, Prestwich et al. (2007) identified a period of 34.4~hr for IC~10 
X-1 and established large radial velocity variations of the He\,{\sc ii} 
$\lambda$4686 line, attributable to orbital motion of the WN star, 
confirming this system as a WR plus BH system.

\begin{figure}[htbp]
%\centerline{\psfig{figure=ngc300x1.eps,width=8cm,angle=0}}
\centerline{\psfig{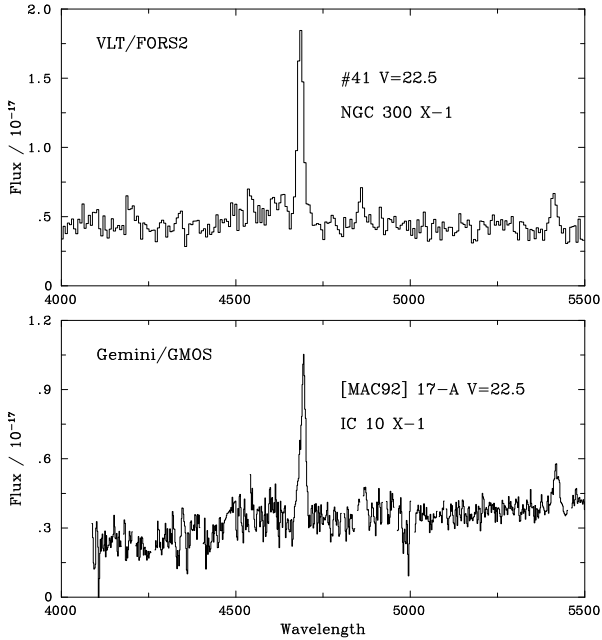}}
\caption{Spectroscopic comparison between NGC~300 \#41, a candidate for 
the  optical counterpart of NGC~300 X-1, and [MAC92] 17-A, alias IC~10 X-1
(Clark \& Crowther 2004; Prestwich et al. 2007).}\label{ngc300x1}
\end{figure}

Carpano et al. (2007a) established that \#41 from Schild
et al. (2003) is the prime candidate for the optical counterpart of 
NGC~300 X-1. 
In Fig.~\ref{ngc300x1} we present VLT/FORS2 spectroscopy
of \#41, which reveals an early-type WN spectrum. A subtype of WN5 results
from the comparable strengths of the 
N\,{\sc v} $\lambda\lambda$4603-20 and N\,{\sc iii} $\lambda\lambda$4634-41
emission lines.

Spectroscopic similarities  between [MAC92] 17-A and NGC~300 \#41,  reinforced
in Fig.~\ref{ngc300x1}, further support a common nature. Indeed, NGC~300 X-1
has X-ray properties reminiscent of IC~10 X-1, including a similar 
period of 32.8~hr (Carpano  et al. 2007b). 
If the early-type WN star \#41 is the optical counterpart to NGC~300 X-1,
what are its properties? MPG/ESO 2.2m Wide Field Imager 
photometry from Carpano (2006) suggests V=22.53
mag and B-V = 0.18 mag, the former in good agreement with V=22.44 
from Schild et al. (2003). For the 1.88~Mpc distance to NGC~300  (Gieren 
et al. 2005), we infer $M_{\rm V} = -5.4$ mag for a reddening of 
E(B-V)=0.5 mag which results from an intrinsic WR colour 
of (B-V)$_0$ = -0.32 mag.

\begin{figure}[htbp]
%\centerline{\psfig{figure=ngc300x1_model.eps,width=8.8cm,angle=-90}}
\centerline{\psfig{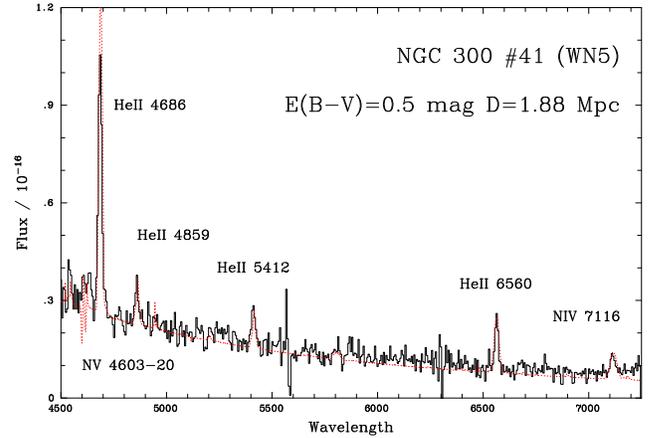}}
\caption{Spectroscopic comparison between the de-reddened VLT/FORS2 spectrum of
NGC~300 \#41 and the theoretical non-LTE model (dotted-lines) discussed in the
text.}\label{ngc300x1_model}
\end{figure}

We have calculated a representative synthetic model using the Hillier 
\& Miller (1998) line-blanketed, non-LTE model atmospheric code.
Relatively simple atomic species are considered in detail, namely H, He,
C, N, O, Ne, Si, S and Fe given the poor quality and limited spectral range of  
the observations. As such, a unique spectral fit may not be claimed, although
a model with the following stellar parameters --
$T_{\ast}$= 65,000\,K, $\log (L/L_{\odot}$)=6.14, $\dot{M}$=7.5$\times 
10^{-6}$
M$_{\odot}$ yr$^{-1}$, $v_{\infty}$=1250 km\,s$^{-1}$ -- reproduces the
observed He\,{\sc ii} $\lambda$4686, 5411, 6560 lines plus
N\,{\sc v} $\lambda\lambda$4603-20 and N\,{\sc iv} $\lambda$7116, 
as shown in Fig.~\ref{ngc300x1_model}. Clumping is accounted for in a crude
manner, with a (maximum) volume filling factor
of 10\%, such that the derived mass-loss rate is equivalent to a
homogeneous mass-loss rate of $\sim 2.5 \times 10^{-5}$ M$_{\odot}$ 
yr$^{-1}$.

For a hydrogen-free WN star, one obtains
a WR mass of 38 M$_{\odot}$ on the basis of the Schaerer 
\& Maeder (1992) mass-luminosity relation. Such a high WR mass together 
with the 32.8~hr period of the
system suggests a companion mass $\geq$ 10 M$_{\odot}$ 
(Carpano et al. 2007b, their fig.~6). This follows from
the need to form an accretion disk around the BH (Ergma \& Yungelson 1998;
Lommen et al. 2005), producing the strong X-ray emission.
Consequently NGC~300 X-1/\#41 shares much in common with 
IC~10 X-1/[MAC92] 17-A for which Clark \& Crowther (2004) obtained
a WR mass of 33\,M$_{\odot}$ using similar techniques, 
and Prestwich et al. (2007) inferred 
a BH mass of 20--30 M$_{\odot}$. 
Consequently, one would expect
radial velocity variations of the WR features of order 100--200 km\,s$^{-1}$, 
both to confirm the WR nature of NGC~300 X-1 and establish a robust BH mass.
Reduced WR and BH masses would follow if  
other sources were to contribute to the ground-based photometry of \#41. 
If the WR star is responsible for 50\% of the measured flux,
its inferred mass would decrease from 38 to 22 M$_{\odot}$, i.e. closer
to the normal WR mass range (Crowther 2007). Consequently, 
our results would greatly benefit from high spatial resolution 
optical imaging.

% Carpano PhD priv. comm..
% B = 22.71
% V = 22.53
% R = 22.03

\section{Summary}

We present new VLT/FORS2 spectroscopy of Wolf-Rayet candidates in the
inner disk of NGC~300 from
Schild et al. (2003), increasing the census of WR stars within the inner disk to 31, 
comprising 16 WN and 15 WC stars. After 
correction for incompleteness, we estimate a true
WR content of $\sim$40, with a subtype ratio close to N(WC)/N(WN)$\sim$0.7.
Curiously, the WC stars in NGC~300 are dominated by early-types, in common with
populations in nearby metal-poor galaxies, yet the H\,{\sc ii} region
measured metallicity of NGC~300 appears relatively metal-rich, with 
$\log$(O/H) + 12 $\sim$8.6. N(WR)/N(O) $\sim$ 0.04 is inferred from the H$\alpha$
derived global star formation rate of NGC~300,
after correction for the $\sim$78\% fraction within the inner disk.

We present spectroscopy of candidate \#41 from Schild et al. (2003) which reveals
an early-type WN5 spectrum, and supports the claim by Carpano et al. (2007ab) that
NGC~300 X-1 is the third known WR plus compact companion system identified. If
\#41 is indeed the optical counterpart of NGC~300 X-1, we estimate WR properties
for which we obtain a mass of 38 M$_{\odot}$. On the basis that an 
accretion
disk is able to form around the compact (BH) component, and so produce strong 
X-ray emission, Carpano et al. (2007b) suggest a companion mass  $\geq$ 
10\,M$_{\odot}$ 
given the 32.8 hr period of the system.

\begin{acknowledgements}
We greatly appreciate the award of ESO Director's Discretionary Time (DDT) which
made the present observations possible. 
\end{acknowledgements}

\end{document}